# Synthesis of monoclinic IrTe$_2$ under high pressure and its physical properties

X. Li[1], J.-Q. Yan[2], D.J. Singh[2], J.B. Goodenough[1], J.-S. Zhou[1]

[1] Materials Science and Engineering program/Mechanical Engineering, University of Texas at Austin, USA

[2] Materials Science and Technology Division, Oak Ridge National Laboratory, Oak Ridge, TN 37831, USA

Abstract

In a pressure-temperature (P-T) diagram for synthesizing IrTe$_2$ compounds, the well-studied trigonal (H) phase with the CdI$_2$-type structure is stable at low pressures. The superconducting cubic (C) phase can be synthesized under higher temperatures and pressures. A rhombohedral phase with the crystal structure similar to the C phase can be made at ambient pressure; but the phase contains a high concentration of Ir deficiency. In this paper, we report that a rarely studied monoclinic (M) phase can be stabilized in narrow ranges of pressure and temperature in this P-T diagram. The peculiar crystal structure of the M-IrTe$_2$ eliminates the tendency to form Ir-Ir dimers found in the H phase. The M phase has been fully characterized by structural determination and measurements of electrical resistivity, thermoelectric power, DC magnetization, and specific heat. These physical properties have been compared with those in the H and C phases of Ir$_{1-x}$Te$_2$. Moreover, magnetic and transport properties and specific heat of the M-IrTe$_2$ can be fully justified by calculations with the density-functional theory presented in this paper.

1. Introduction

The 5d transition metal dichalcogenide $IrTe_2$ has attracted significant interest due to the close interplay between structural instability, charge/orbital density wave (CDW/ODW), and superconductivity at the presence of a strong spin-orbit coupling (SOC).[1-7] If synthesized under ambient pressure, the stoichiometric $IrTe_2$ crystallizes in a trigonal (H) phase with the $CdI_2$-type crystal structure at room temperature.[8] Upon cooling, the $IrTe_2$ undergoes a structural transition at 280 K from the high temperature H phase (space group *P-3m1*) to a low temperature triclinic phase (space group *P-1*).[9] This structural transition can be suppressed by either partially substituting Ir with other transition metal ions (e.g. Pt, Pd)[6, 7] or intercalating Cu into the layered $IrTe_2$ structure.[10] As the doping content increases, the structural transition is suppressed and disappears completely where superconductivity emerges. The competition between the structural transition and superconductivity has also been confirmed by applying hydrostatic pressure on a doped superconducting $IrTe_2$;[2] pressure stabilizes the low-temperature triclinic phase and suppresses superconductivity. A variety of mechanisms have been proposed to rationalize the trigonal to triclinic phase transition, such as the formation of a charge-orbital density wave (CDW/ODW),[7] the bonding instabilities associated with the Te-5p states[11], and the formation of Ir-Ir dimers.[9] Nevertheless, there is no doubt that an electron-lattice interaction is very strong in the H phase of $IrTe_2$.

Motivated to study the structure-property relationship in $IrTe_2$, we have synthesized an alternative phase of $IrTe_2$ under high pressure. The H phase of $IrTe_2$ becomes unstable if the synthesis is performed under high pressure. As the pressure used in the synthesis increases, a monoclinic (M) phase has been found at 5 GPa and subsequently a cubic phase at 20 GPa.[12, 13] It should be noted that an $Ir_3Te_8$ with rhombohedral symmetry and a high concentration of Ir deficiency in a general formula $Ir_{1-x}Te_2$ can be synthesized under ambient pressure.[14] All three crystal structures of $Ir_{1-x}Te_2$ can be characterized by different stacking sequences of layers of edge-shared or corner-shared $IrTe_6$ octahedra. As shown in Fig.1, the H-$IrTe_2$ is composed of stacking layers of edge-shared IrTe$_6$ octahedra along the *c* axis; Ir ions are connected across shared octahedral-site edges to become a regular close-packed triangular lattice projected along the c axis (Fig. 1a). The cubic Ir$_{1-x}$Te$_2$ (space group *Pa-3*) has a 3D framework of corner-shared octahedra (Fig.1b). We define the 3D structure consisting of corner-shared octahedra as the C phase throughout the paper. The M-IrTe$_2$ has segments of the H phase connected by corner-shared octahedral (Fig.1c). It is clear that the pressure effect on synthesizing IrTe$_2$ is to increase the population of corner-shared octahedral with a decrease of Ir deficiency.

Physical properties of both H and C phases of IrTe$_2$ have been well studied.[1-9, 11-25] The C-Ir$_{1-x}$Te$_2$ is a superconductor with T$_c$ =1.8 - 4.7 K depending on the Ir nonstoichiometry x. It has been shown that the C-Ir$_{1-x}$Te$_2$ with Ir vacancies can be synthesized under a broad range of pressure.[14, 24] An octahedral-site distortion becomes more severe as pressure increases. Since the Fermi level cuts into to the band primarily from Te-5p and Ir-4d orbitals in the C-Ir$_{1-x}$Te$_2$, transport properties would be significantly influenced by Ir vacancies. For a stoichiometric C-IrTe$_2$, a broader conduction band gives rise to metallic conductivity.[24] Previous experimental and theoretical studies on either the C or the H phase of IrTe$_2$ have indicated a strong electron-lattice coupling. In contrast, as far as we know, only the material synthesis and the structural information of the M-IrTe$_2$ have been reported. It is important to complete the study on all three phases of IrTe$_2$ and to make their comparison.

In this paper, we report the phase diagram of IrTe$_2$ as the material is synthesized under high temperature and high pressure. The M-IrTe$_2$ is stabilized within narrow ranges of pressure and temperature. Measurements of electrical resistivity and thermoelectric power reveal that the M-IrTe$_2$ is a non-Fermi liquid metal. In addition, specific heat and DC magnetic susceptibility data indicate that the M-IrTe$_2$ is diamagnetic with no phase transition from room temperature down to 1.8 K. We also investigated the crystal structure of M-IrTe$_2$ by *in-situ* X-ray powder diffraction under high pressure. Observed physical properties of the M-IrTe$_2$ are consistent with predictions from density-functional theory (DFT) calculations.

2. Experimental details

The H-IrTe$_2$ phase was prepared by firing a stoichiometric mixture of Ir (Alfa, 99.99%) and Te (Alfa, 99.999%) powders at 1000 °C for 2 days sealed in an evacuated quartz tube. The M-IrTe$_2$ phase in the present study was synthesized under high pressure and high temperature (HPHT) with a Walker-type multi-anvil module (Rockland Research Co.). The starting material of the H-IrTe$_2$ or a mixture of Ir and Te powders with a ratio of 1:2 were pressed into a small pellet and loaded in a BN capsule, which prevents any direct contact between the Pt heater and the sample. The capsule together with two LaCrO$_3$ end disks were placed inside of a Pt heater, which was inserted into a MgO octahedron with a LaCrO$_3$ sleeve. The whole assembly was kept under a high pressure of 3 - 5.5 GPa and high temperature of 650-1400 °C for several hours before quenching to room temperature.

The phase purity of the high-pressure products was examined by powder X-ray diffraction (XRD) at room temperature with a Philips X'pert diffractometer (Cu Kα radiation). The crystal structure of M-IrTe$_2$ under pressure was studied with a diamond anvil cell (DAC). The sample was loaded in the DAC with a small amount of Au powder as the pressure manometer. Lattice parameters and atom positions were obtained by refining the XRD patterns with the software FULLPROF. Resistivity and specific-heat data were collected with a Physical Property Measurement System (PPMS), Quantum Design (QD). DC magnetization was measured with a superconducting quantum interference device (SQUID) magnetometer, QD. Thermoelectric power measurements were performed in a homemade setup. DFT calculations were conducted with the general potential LAPW method as implemented in the WIEN2k code.[26] We used the generalized gradient approximation of Perdew, Burke and Ernzerhof (PBE-GGA)[27] with well converged basis sets and zone samplings. The conductivity anisotropy and thermoelectric power were calculated within Boltzmann transport theory by using the BoltzTraP code.[28]

3. Results and discussion

We have used the H-IrTe$_2$ or a mixture of Ir and Te powder in 1:2 ratio as the starting material and performed the mapping out of products in the pressure-temperature diagram of Fig. 2. The M-IrTe$_2$ was stabilized in a small area of the diagram; the detailed information of these high pressure products is listed in Table 1. By sintering at 1000 °C and ambient pressure, the H-IrTe$_2$ was obtained as a black powder with a trigonal symmetry as reported previously.[8] Within a pressure range of 3.5 ≤ P ≤ 4.5 GPa and at T < 800 °C, a nearly pure M-IrTe$_2$ phase was obtained. At P = 4.5 GPa, we found a structural evolution from the H phase (if the H phase is used as the starting material) to the M phase and then to the C phase as the sintering temperature increased. The H phase remains stable up to 650 °C; it transforms into the M phase at T > 700 °C. However, high-pressure products always show a two-phase coexistence of the M and the C phases once the H phase disappears at T > 700 °C. The volume fraction of the C phase increases as the sintering temperature further increases. We have obtained a pure C-IrTe$_2$ phase at T=1300 °C. It is obvious that the synthesis at higher pressures and temperatures favor the C phase. This observation is consistent with a report in the literature.[24] The C-phase samples with a high concentration of Ir deficiency can be synthesized under ambient pressure.[14] The Ir deficiency reduces as the synthesis pressure increases. The C phase obtained in this work shows the lowest Ir deficiencies as far as we know. This assertion will be further elaborated below where we present the physical properties.

Fig. 3 shows the result of Rietveld refinement of the XRD pattern of the M-IrTe$_2$ synthesized at 4.5 GPa and 740 °C. The XRD pattern shows the two-phase coexistence. The majority phase can be indexed with a monoclinic unit cell (space group *C2/m*, No. 12) with lattice parameters of $a$ = 19.9455(6) Å, $b$ = 3.9964(1) Å, $c$ = 5.3133(2) Å, and $\beta$ = 90.771(2)°, which agrees well with the M phase reported in the literature.[12] The impurity phase can be indexed and refined with the C phase structural model. The refinement indicates ~2.0 wt.% of the C-IrTe$_2$ in this sample. The refined atomic parameters of the M-IrTe$_2$ shown in Table 2, are also consistent with predictions from the DFT calculation presented in this work. The normalized cell volume (per formula unit) V = 70.58 Å$^3$ of the M-IrTe$_2$ is between that of H-IrTe$_2$ (72.22 Å$^3$) and that of C-IrTe$_2$ (66.25 Å$^3$), which is consistent with the structural change with increasing pressure from the layered phase with only edge-shared octahedra in the H phase to a mixture of edge-shared and corner-shared octahedra in the M phase and then to the 3D structure with corner-shared octahedra only in the C phase.

Li et al.[25] have calculated the equation of state for all possible polymorphs with the IrTe$_2$ formula. However, as far as we know, no experimental structural study under pressure has been performed for the M-IrTe$_2$. We have carried out an in-situ high pressure structural study up to 8 GPa by using a DAC at room temperature. The refinement was made on the XRD patterns for P ≤ 6 GPa only. The quality of XRD patterns taken at P > 6 GPa deteriorates too much for a sound refinement; but no phase transition can be identified. Fig. 4 shows lattice parameters of the M phase under pressure. All parameters except $\beta$ decrease monotonically with increasing pressure; an increase of $\beta$ indicates that pressure enlarges the structural distortion. The crystal structure of the M phase remains stable to the highest pressure in this study, which is in line with the calculation by Li et al.[25] The bulk modulus $B_0$ = 95 ± 16 GPa of the M-IrTe$_2$ was obtained by fitting the V versus P plot with the Birch-Murnaghan equation,[29] which is comparable to 132(9) GPa in the H phase and 126(5) GPa in the C phase[13]. The difference of bulk modulus between the M phase and the C phase is just slightly larger than the experimental uncertainty. More experiments are needed to verify these results. One may be curious why a 3D structure of the C phase has a lower bulk modulus than that of the layered H phase. In order to answer this question, we have to take a close look at the local structures in these phases. IrTe$_6$ octahedra are basic units for all three H, M, and C phases. As shown in Table 3, the averaged Ir-Te bond length in an octahedron is 2.639 Å in the H phase, which is smaller than that (2.657 Å) in the C phase. Whereas the C phase has a 3D structure, the Ir-Te bonds of the C phase carry more ionic character than those in the H structure. Given the relatively larger experimental uncertainty in the structural determination of the M phase, we have to take the optimized Ir-Te bond length from a

DFT calculation for comparison; the averaged bond length of the M phase is indeed extremely close to that of the C phase. It appears that the bulk modulus in the $IrTe_2$ phases is determined by the local bonding character instead of how the octahedra are packed in a structure. This conclusion is consistent with the presence of interlayer bonding in the layered H phase and strong Te-Te interactions associated with p-electron bonding. In any case, the H phase clearly transforms into the C phase under high pressure.

Fig. 5 shows the temperature dependence of electrical resistivity $\rho(T)$ and thermoelectric power $S(T)$ of M-$IrTe_2$ from 1.8 to 300 K. $\rho(T)$ exhibits a linear temperature dependence from 100 K to 400 K, the highest temperature in this work. A similar $\rho(T)$ has been reported in the C-$Ir_3Te_8$ from 20 to 700 K excluding a narrow temperature range at $T_s$.[14] The first-order transition in the H phase leads to an obvious anomaly in $\rho(T)$ near 280 K. However, a nearly linear $\rho(T)$ from 30 K to 200 K is still visible.[11] Therefore, a linear $\rho(T)$ is a common feature for the three different phases of $Ir_{1-x}Te_2$. In a plot of $\rho$ versus $T^2$ as an inset of Fig. 5, the M phase does not show a Fermi-liquid behavior to the lowest temperature. In fact, a shallow minimum of $\rho(T)$ is nearly visible, indicating a possible Kondo effect. In addition, a sharp drop in the $\rho(T)$ was observed at ~3K. The onset temperature of the drop shifts to lower temperatures under external magnetic fields (the top inset of Fig. 5a) and the resistivity does not reach zero at 1.8 K, which indicates filamentary superconductivity. Considering that our M-$IrTe_2$ samples have a small fraction of C-$IrTe_2$ impurity and the C phase becomes a superconductor below 3K, [24] we could attribute filamentary superconductivity to the C-$IrTe_2$ impurity phase in the sample. One may question that the overall $\rho(T)$ of our M-phase samples could be dominated by the presence of C phase as the impurity. This concern can be eased by the observation that the $\rho$ of the C phase is about five times higher than that of the M phase. The metallic M phase can also be characterized by the linear relationship of S verse T in Fig. 5b as predicted from the Mott diffusive formula.

Fig. 6 shows temperature dependence of magnetic susceptibility $\chi(T)$ measured with zero-field cooling (ZFC) and field cooling (FC) from 1.8 to 300 K in an applied magnetic field of H = 0.5T. χ(T) curves of ZFC and FC overlap and show a temperature-independent diamagnetism for T > 15 K. As a matter of fact, all three phases of $Ir_{1-x}Te_2$ show a diamagnetism due to the dominant contribution from the diamagnetism of the core electrons.[11, 14] The diamagnetism is offset by a smaller positive value of the Pauli paramagnetism from electrons near the Fermi energy. χ(300K) for all three phases of $Ir_{1-x}Te_2$ are listed in Table 3. Based on the tabulated value of diamagnetism from core electrons, we were able to calculate the Pauli paramagnetism for all three phases, which together with γ from specific heat allow us to infer values for an effective paramagnetic Wilson ratio $R_w = \gamma/\chi_0$ for the conduction electrons of these phases as shown in the Table 3. It should be kept in mind that these compounds are far from ordered magnetism, which reflects the importance of Te-p states in the electronic structure near the Fermi energy.

Fig. 7 shows the specific heat $C_p(T)$ data of $M-IrTe_2$ in the temperature range from 0.05 to 300 K in magnetic fields of H =0, 5T and 9T. No obvious field effect was observed in the plot over a broad temperature range. However, a plot of $C_p/T$ verse $T^2$ (the inset of Fig. 7) at extremely low temperatures shows clearly the field dependence. Fitting with the formula for specific heat $C_P/T = \gamma + \beta T^2$ gives a field independent γ = 1.8(1) mJ/(mol $K^2$) and a β = 0.42(4) mJ/(mol $K^4$) and therefore the Debye temperature $\theta_D$ = 240(8) K from the curve with H=0. However, a field dependent β does not make sense if the $T^3$ term in $C_p$ is solely from the lattice contribution. The unusual field dependence of $C_p$ due to electrons or magnetic excitations deserves further study.

By using $H-IrTe_2$ as the starting material, we were also able to synthesize single-phase $C-IrTe_2$ at P ≥ 4.5 GPa and 1300 °C. None of the C phases in the literature is chemically stoichiometry. The $C-Ir_{1-x}Te_2$ synthesized under ambient pressure has the highest Ir deficiency with x=0.25.[14] The Ir deficiency x decreases as the synthesis pressure increases. The C phase with Ir deficiency is a superconductor; $T_c$ increases monotonically with decreasing x. A plot of $T_c$ versus 1-x from samples in the literature is shown in Fig. 8(a).[24] Although there is a symmetry change from the rhombohedral phase of $Ir_{0.75}Te_2$ synthesized under ambient pressure to the cubic phase of $Ir_{1-x}Te_2$ ( $0.05 < x < 0.15$) synthesized under high pressure, data of $T_c$ versus 1-x can be fit linearly. The C-$Ir_{1-x}Te_2$ synthesized in this work shows the highest $T_c$ as far as we know. By extrapolating the fitting line, we can estimate an x=0.03-0.04 for our C-$Ir_{1-x}Te_2$ sample. The behavior of resistivity drop under different magnetic fields and a clear anomaly in the $C_p$ at low temperatures confirm bulk superconductivity in the C-phase sample. As shown in Fig. 8(b), the electronic contribution $\gamma$ to the $C_p$ was extracted by suppressing superconductivity under a magnetic field H= 8 T.

Table 3 illustrates local structures and physical properties of $Ir_{1-x}Te_2$ in three different phases. The Wilson ratio $R_w$=1 holds for free electrons. A larger $R_w$ for all three $Ir_{1-x}Te_2$ indicates a correlation-enhanced magnetism. A slightly higher $R_w$ =3.03 for the M phase than those for the H and C phases is still within the measurement uncertainty since a relatively larger error bar is expected in determining a much smaller $\gamma$ for the M phase. The data in Table 3 make easier a discussion to correlate electronic properties with the structural evolution. The trigonal arrangement of Ir inside a layer for the H phase makes highly degenerate electronic states unstable against the dimerization below $T_s$; the transition opens up a gap in $d_{xy}$ and $d_{yz}$ bands associated with orbitals on Ir, but leaves the $d_{zx}$ unchanged.[22] In the M phase, however, only the structural segment of two edge-shared octahedra rows remains. The Ir-Ir distances in an Ir triangle in edge-shared octahedra rows are no longer identical, which eliminates the band degeneracy and therefore the structural instability for the dimerization. The 3D structure of the C phase creates a higher band degeneracy as seen from a relatively higher $\gamma$. However, it appears to cost a much higher elastic energy with a major structural reconstruction to form dimmers in this 3D structure.

We calculated the electronic structure of the M-$IrTe_2$ within density-functional theory (DFT) as shown in Fig. 9. The optimized crystal structure is close to the experimental result; the atomic positions and local bond lengths and bond angles are listed in Table 2 and 3 for a side-by-side comparison. The density of states (DOS) shows that the N($E_F$) is 0.54 eV$^{-1}$ per IrTe$_2$ formula unit, in which Ir-4d electron contribution is only about 0.18 eV$^{-1}$ (Fig. 9a). The Te-p states dominate the electronic structure near the Fermi energy $E_F$, N($E_F$). The very low N($E_F$) indicates that the M-IrTe$_2$ is far from a ferromagnetic instability. The electronic contribution to the specific heat was calculated as $\gamma_0$ = 1.26 mJ/(mol K$^2$). An enhancement factor $\lambda$ = 0.4 can be derived from the equation $\gamma = (1+\lambda)\gamma_0$ based on the experimental result $\gamma$ = 1.8 mJ/(mol K$^2$). Moreover, as a result of the structural distortion in M-IrTe$_2$, some modest nesting features associated with flat parts of Fermi surface predict that the M-IrTe$_2$ is an anisotropic metal (Fig. 9b). The conductivity anisotropy was calculated by using the expression for σ/τ, where σ is the conductivity and τ is a scattering time, which is assumed isotropic. The calculation gives a highly anisotropic conductivity in the M phase, which remains to be verified by measuring a single crystal sample of the M phase. The calculated thermoelectric power is also anisotropic with a large negative value in the lower conduction direction (-20.4μV/K at 300 K) along the *c* axis but a small value in higher conduction direction (0.9μV/K at 300 K) within the *ab* plane. The total thermoelectric power S = -3.3 μV/K for the polycrystalline sample is obtained with the formula S = (σ$_x$S$_x$+σ$_y$S$_y$+σ$_z$S$_z$)/(σ$_x$+σ$_y$+σ$_z$), which matches the experimental result remarkably well.

4. Conclusion

The monoclinic phase of IrTe$_2$, an alternative crystal structure to the well-known trigonal and cubic phases, can be stabilized in narrow ranges of pressure and temperature in the P-T diagram. Like the H phase, the M-IrTe$_2$ is a diamagnetic metal to the lowest temperature. The monoclinic structure contains blocks of the edge-shared IrTe$_6$ octahedra similar to the H phase. However, the intrinsic structural distortion of the M phase lifts the degeneracy found in the H phase. As a result, the metallic M phase remains stable to the lowest temperature. The superconducting C-Ir$_{1-x}$Te has been obtained by syntheses in broader ranges of temperature and pressure. T$_c$ increases as the Ir deficiency decreases. We have obtained a C-Ir$_{0.97}$Te$_2$ sample with $T_c \approx 5.4$ K synthesized under 4.5 GPa and 1300 °C. A smaller $\gamma = 9$ mJ/(mol K$^2$) for the C-Ir$_{0.97}$Te$_2$ than that $\gamma = 11$ mJ/(mol K$^2$) for Ir$_3$Te$_8$ made under ambient pressure indicates that the Ir vacancies instead of the density of state entering in the BCS expression of $T_c$ plays a dominant role to determine the transition temperature in this possible BCS superconductor. The band structure and physical properties of the M phase have been calculated with the density functional theory. The calculation gives highly anisotropic transport properties. While it is difficult to confirm them with a polycrystalline sample, the overall conductivity and thermoelectric power for polycrystalline samples match well the experimental results.

**Acknowledgment**

This work was supported by the National Science Foundation (DMR MIRT 1122603). Work at ORNL was supported by the US Department of Energy, Office of Science, Basic Energy Science, Materials Sciences and Engineering Division.

jszhou@mail.utexas.edu

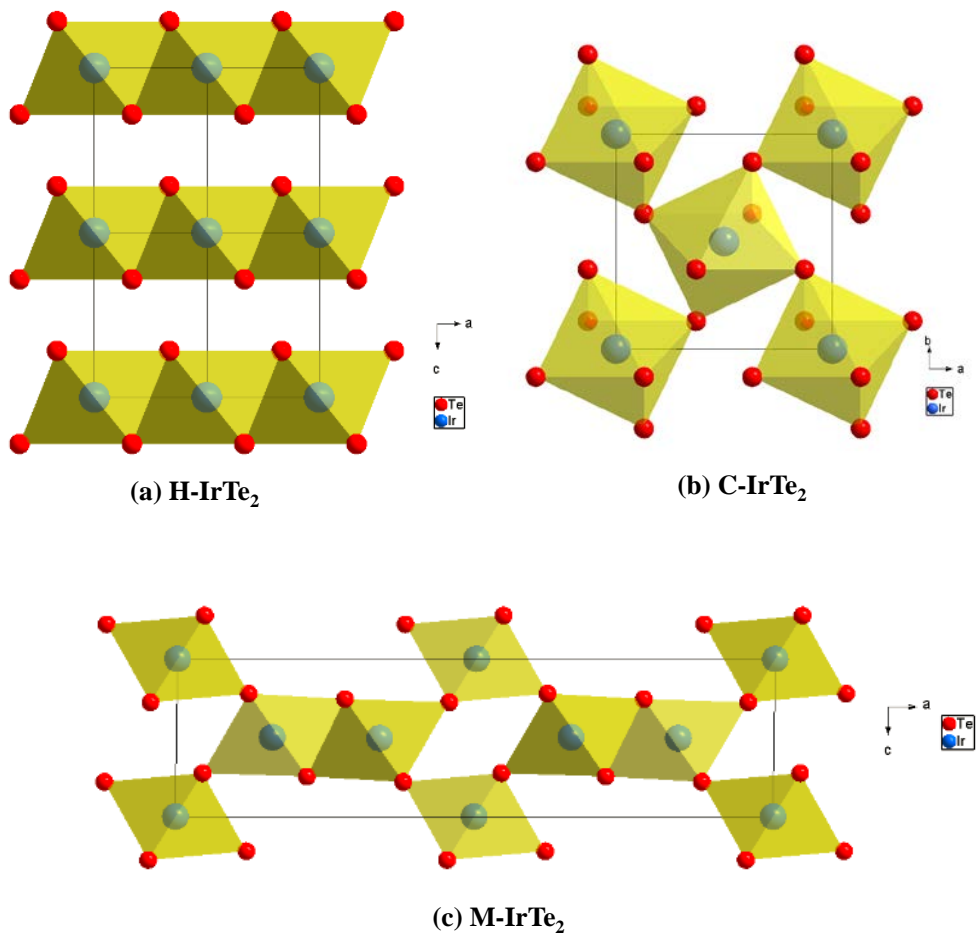

**Fig. 1** (Color online) Structures of different IrTe$_2$ phases. (a) the H phase of IrTe$_2$. (b) the C phase of IrTe$_2$. (c) the M phase of IrTe$_2$.

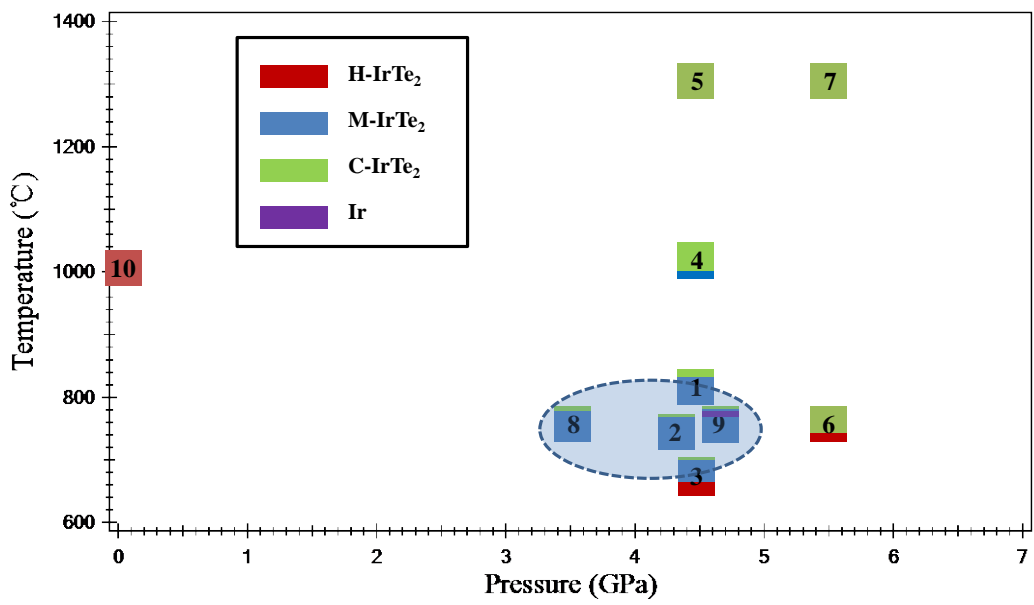

**Fig. 2** The pressure-temperature diagram for synthesizing a variety of phases of IrTe$_2$.

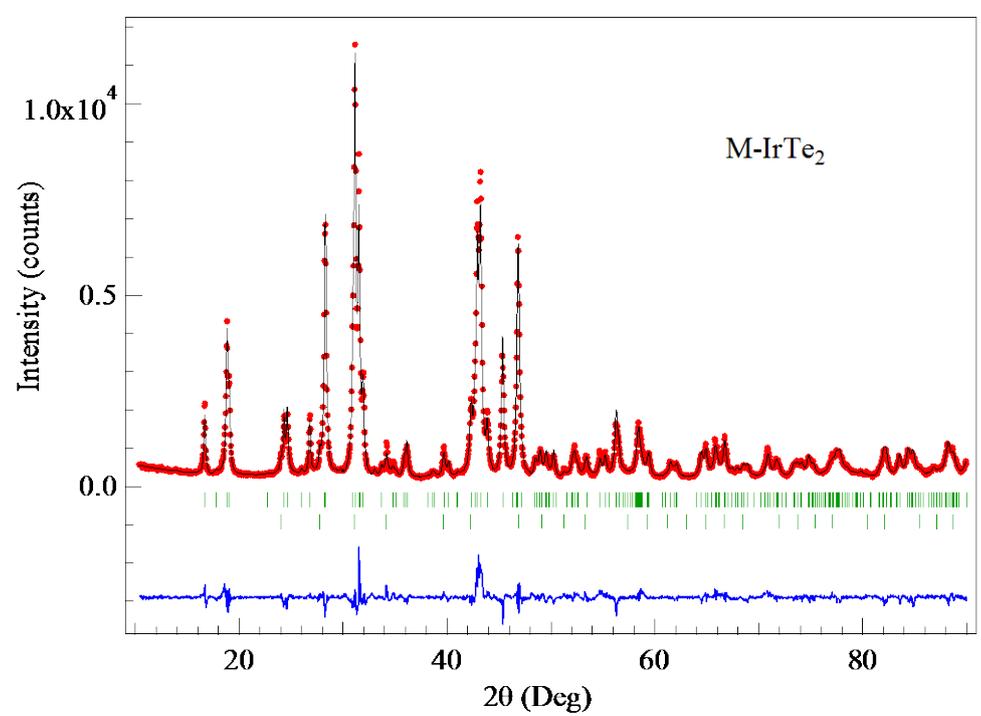

**Fig. 3** (Color online)　X-ray powder diffraction pattern of the M-IrTe$_2$ and the result of Rietveld refinement.

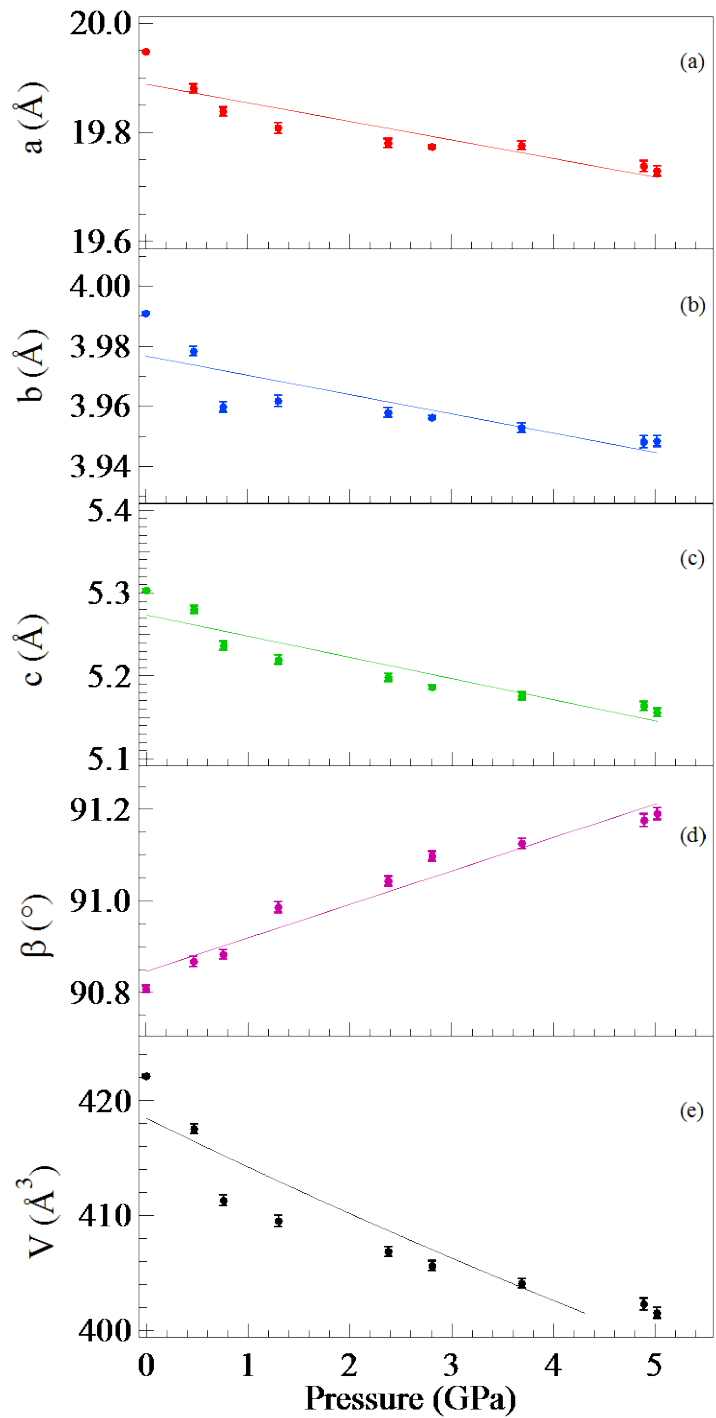

**Fig. 4** (Color online) (a) – (d) Pressure dependences of lattice parameters of the M-IrTe$_2$. (e) the plot of V versus P and the fitting curve with the Birch-Murnaghan equation.

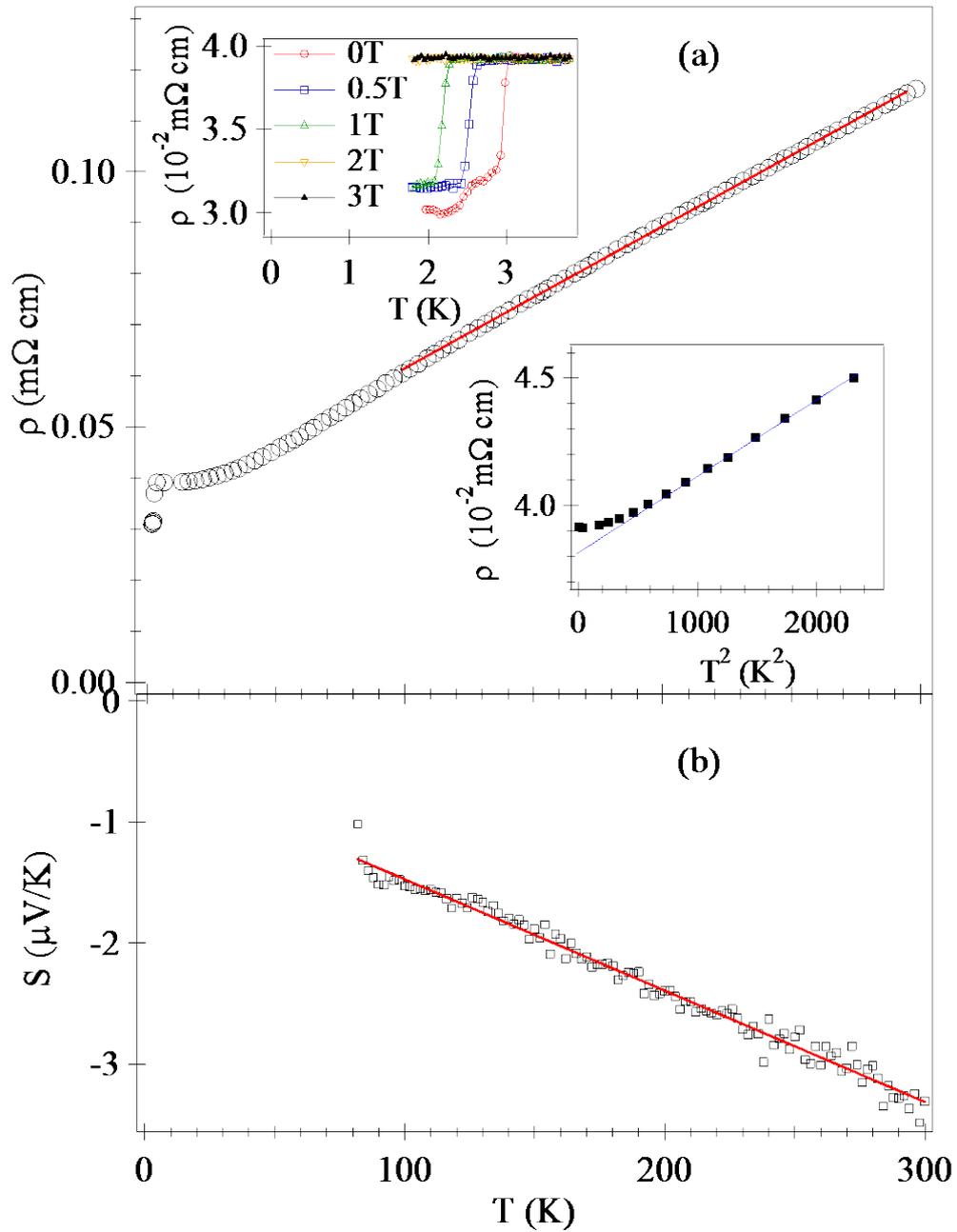

**Fig. 5** (a) (Color online) Temperature dependence of the resistivity $\rho$(T) for the M-IrTe$_2$; insets (upper) a zoom-in plot of $\rho$(T) under different fields; (lower) a plot of $\rho$ versus T$^2$. (b) Temperature dependence of the thermoelectric power $S$ (T) for the M-IrTe$_2$.

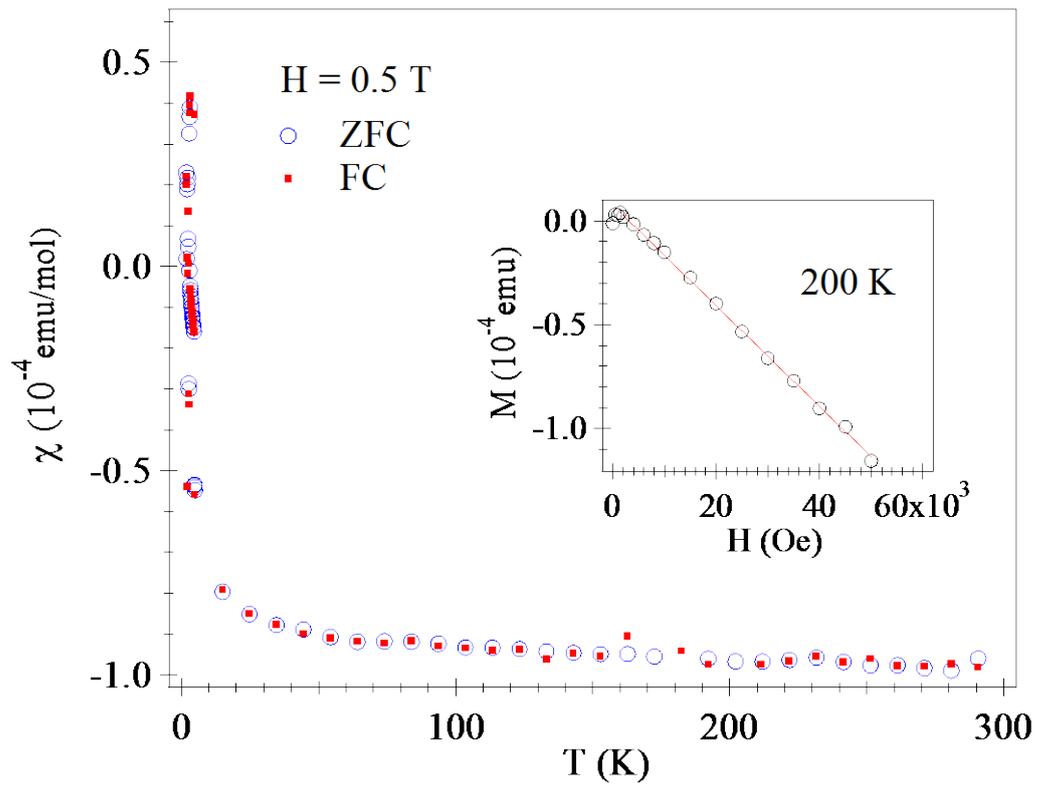

**Fig. 6** (Color online) Temperature dependence of magnetic susceptibility χ (T) of the M-IrTe$_2$ measured from 1.8 to 300 K under H = 0.5T after zero-field cooling (ZFC) and field cooling (FC); inset: the field dependence of magnetization M at 200 K.

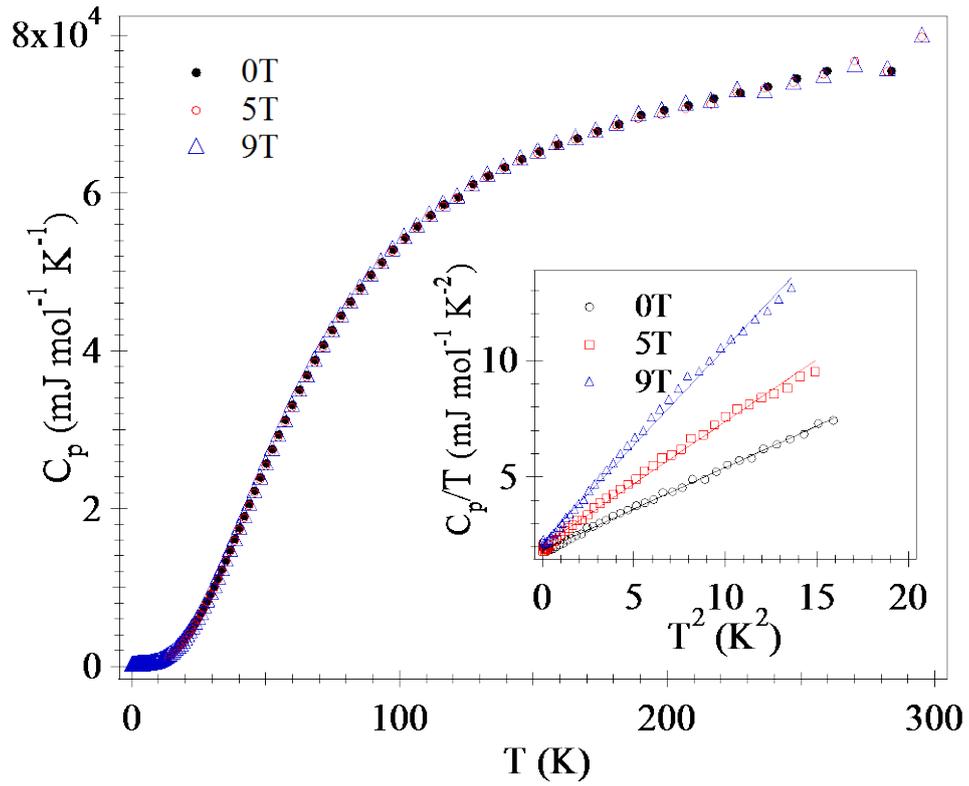

**Fig. 7** (Color online) Temperature dependence of specific heat of the M-IrTe$_2$ measured from 0.05 to 300 K under H = 0T, 5T, 9T; inset: a plot of $C_p/T$ vs $T^2$ at low temperatures under different magnetic fields.

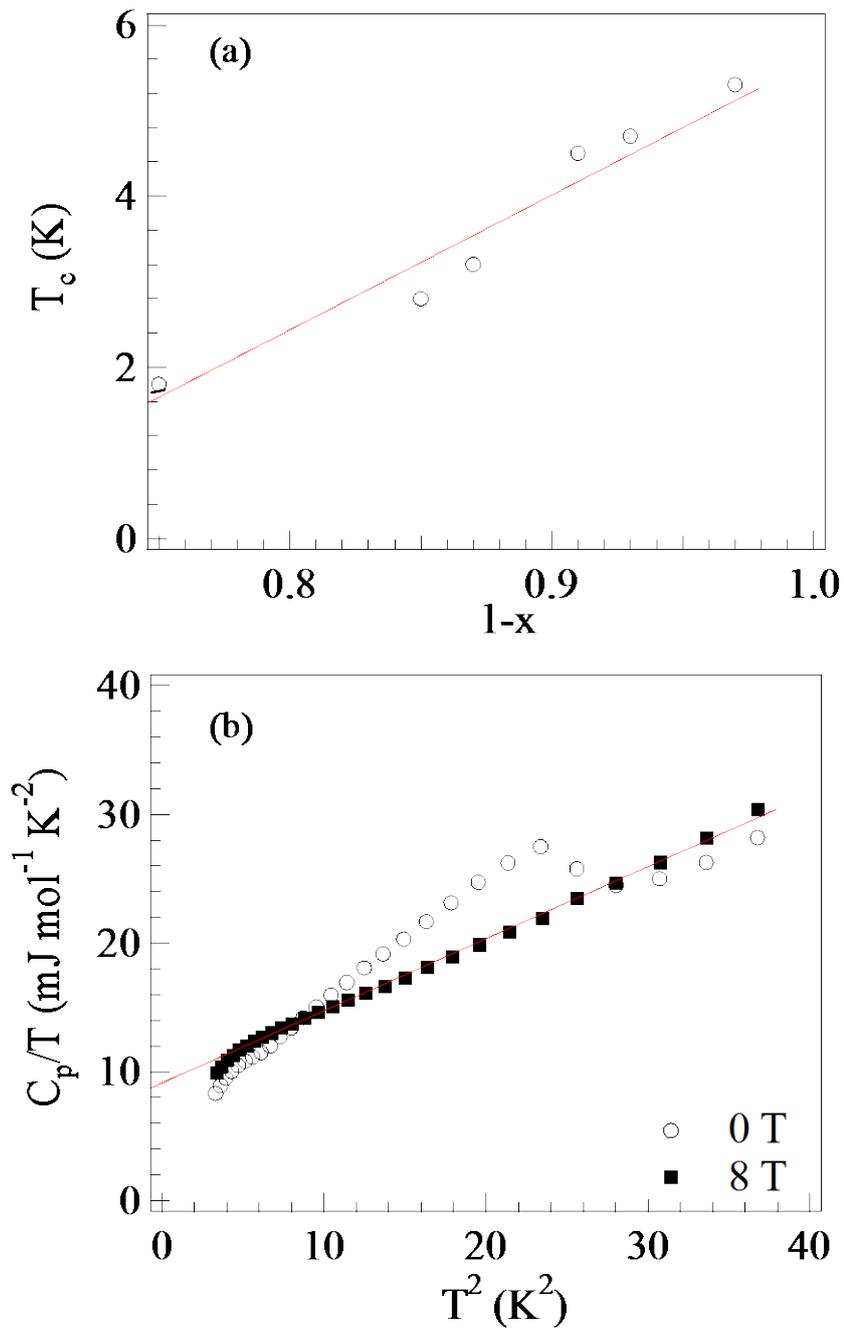

**Fig. 8** (Color online) (a) $T_c$ vs Ir concentration 1-x for the C-IrTe$_2$. (b) $C_p/T$ vs $T^2$ at low temperature for the C-Ir$_{0.97}$Te$_2$.

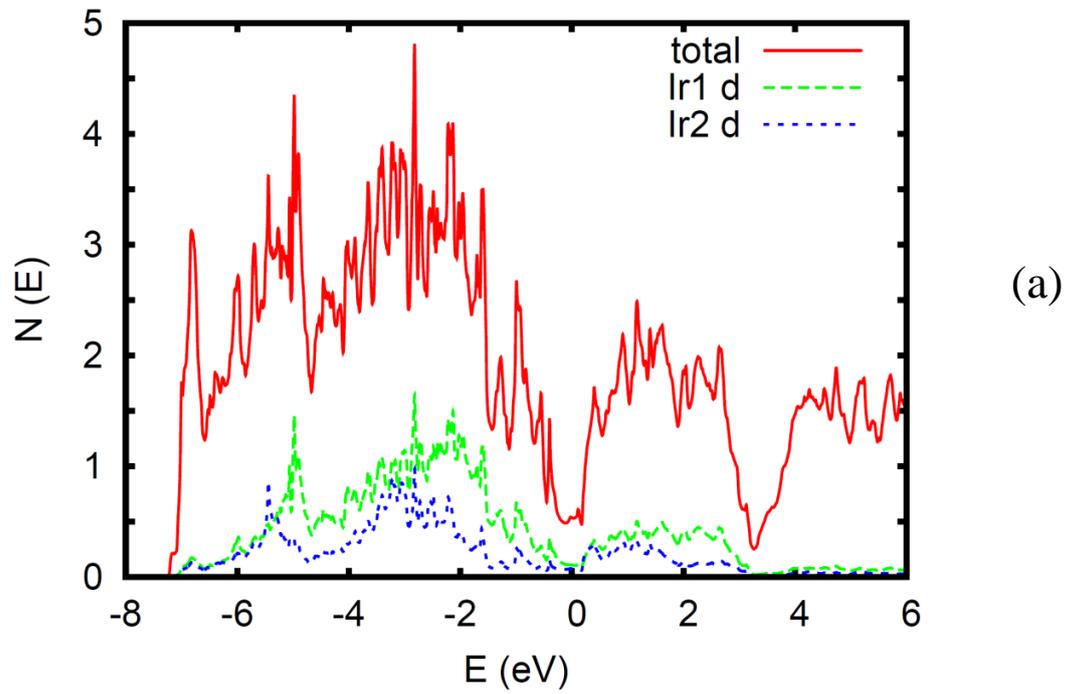

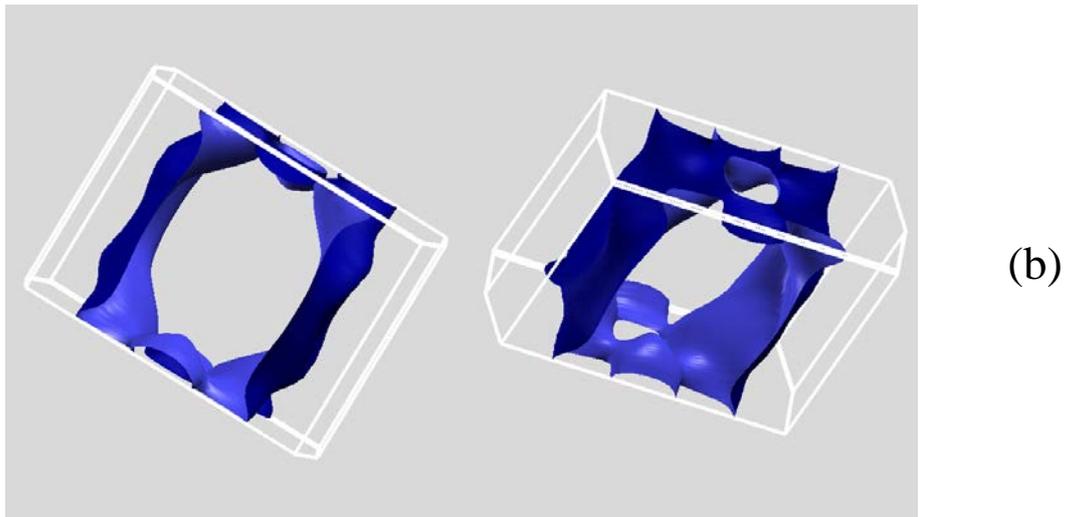

**Fig. 9** (Color online) DFT calculations of electronic structure for the M-IrTe$_2$. (a) Density of states (DOS). (b) Fermi surfaces with nesting features.

**Table 1**. A summary of the IrTe$_2$ products synthesized under HPHT conditions.

| Sample | Starting material | Pressure (GPa) | Temperature (°C) | Time (hour) | Phases |
|---|---|---|---|---|---|
| 1 | H-IrTe$_2$ | 4.5 | 770 | 4.5 | M-IrTe$_2$ (94%) + C-IrTe$_2$ (6%) |
| 2 | H-IrTe$_2$ | 4.5 | 740 | 4.5 | M-IrTe$_2$ (97%) + C-IrTe$_2$ (3%) |
| 3 | H-IrTe$_2$ | 4.5 | 680 | 4.5 | M-IrTe$_2$ (63%)+ H-IrTe$_2$ (29%)+ C-IrTe$_2$ (2%) |
| 4 | Ir+Te | 4.5 | 1000 | 4.5 | M-IrTe$_2$ (10%) + C-IrTe$_2$ (90%) |
| 5 | H-IrTe$_2$ | 4.5 | 1300 | 4.5 | C-IrTe$_2$ |
| 6 | H-IrTe$_2$ | 5.5 | 750 | 4.5 | H-IrTe$_2$ (10%)+ C-IrTe$_2$ (90%) |
| 7 | H-IrTe$_2$ | 5.5 | 1300 | 4.5 | C-IrTe$_2$ |
| 8 | H-IrTe$_2$ | 3.5 | 750 | 4.5 | M-IrTe$_2$ (93%) + C-IrTe$_2$ (7%) |
| 9 | H-IrTe$_2$ +Ir (2%) | 4.5 | 750 | 0.5 | M-IrTe$_2$ (94%) + C-IrTe$_2$ (4%) +Ir (2%) |
| 10 | Ir+Te | 0 | 1000 | 48 | H-IrTe$_2$ |

**Table 2**. Atomic coordinates and isotropic thermal factors for the M-IrTe$_2$ from X-ray powder diffraction data[a] and calculations.

| Atom | Site | x (exp.) | x (cal.) | y (exp.) | y (cal.) | z (exp.) | z (cal.) | B (Å$^2$) |
|---|---|---|---|---|---|---|---|---|
| Ir1 | 4i | 0.3397(2) | 0.3401 | 0 | 0 | 0.5024(8) | 0.5 | 0.6(1) |
| Ir2 | 2a | 0 | 0 | 0 | 0 | 0 | 0 | 0.3(1) |
| Te1 | 4i | 0.4560(3) | 0.4556 | 0 | 0 | 0.2760(9) | 0.2792 | 0.5(1) |
| Te2 | 4i | 0.2193(3) | 0.2205 | 0 | 0 | 0.7395(9) | 0.7477 | 0.6(1) |
| Te3 | 4i | 0.1199(2) | 0.1217 | 0 | 0 | 0.2202(8) | 0.2106 | 0.6(1) |

[a] Discrepancy factors: Rp = 6.64%, Rwp = 8.53%, Rexp = 3.53%, $\chi^2$ = 5.82, $R_{Bragg-factor}$ = 4.67, $R_{f-factor}$ = 2.73.

Space group $C2/m$ (No. 12), a = 19.9455(6) Å, b = 3.9964(1) Å, c = 5.3133(2) Å, β = 90.771(2)°, V=423.48(2) Å$^3$, Z=6.

**Table 3.** Local structures and physical properties of Ir$_{1-x}$Te$_2$ in three different phases.

| Phase | | H-IrTe$_2$ | M-IrTe$_2$ | | | | | C-IrTe$_2$ |
|---|---|---|---|---|---|---|---|---|
| Space group | | *P -3 m 1* (No. 164) | *C2/m* (No. 12) | | | | | *P a -3* (No. 205) |
| Lattice parameters | | a=3.9280Å, c=5.4050Å, Z=1, V/Z=72.22Å$^3$ [13] | a=19.9455Å, b=3.9964Å, c=5.3133Å, β = 90.771°, Z=6, V/Z=70.58Å$^3$ | | | | | a=6.4320Å, Z=4, V/Z=66.52Å$^3$ [13] |
| | | | EXP | | | DFT | | |
| Intralayer | Ir-Te (Å) | 2.6399 | Ir1 | -Te1 | 2.6273 | Ir1 | -Te1 | 2.5993 | 2.6578 |
| | | | | -Te2 | 2.6436×2 | | -Te2 | 2.6727×2 | |
| | | | | -Te2 | 2.7268 | | -Te2 | 2.7399 | |
| | | | | -Te3 | 2.6043×2 | | -Te3 | 2.6280×2 | |
| | | | <Ir1-Te> | | 2.6417 | <Ir1-Te> | | 2.6567 | |
| | | | Ir2 | -Te1 | 2.6359×4 | Ir2 | -Te1 | 2.6482×4 | |
| | | | | -Te3 | 2.6481×2 | | -Te3 | 2.6592×2 | |
| | | | <Ir2-Te> | | 2.6400 | <Ir2-Te> | | 2.6519 | |
| | Ir-Ir (Å) | 3.9280 (T > T$_s$); 3.119 (T < T$_s$) 3.905-4.030 (T < T$_s$) [9] | Ir1-Ir1 | | 3.9964 4.0981×2 | Ir1-Ir1 | | 3.9964 4.1123×2 | 4.5481 |
| | <Te-Te> (Å) | 3.5280 | Ir1 octahedron | | 3.7010 | Ir1 octahedron | | 3.7426 | 3.7541 |
| | | | Ir2 octahedron | | 3.7289 | Ir2 octahedron | | 3.7456 | |
| | Te-Ir-Te (°) | 180 | Te1-Ir1-Te2 | | 179.732 | Te1-Ir1-Te2 | | 178.123 | 180 |
| | | | Te2-Ir1-Te3 | | 170.894 | Te2-Ir1-Te3 | | 169.188 | |
| | | | Te1-Ir2-Te1 | | 180 | Te1-Ir2-Te1 | | 180 | |
| | | | Te3-Ir2-Te3 | | 180 | Te3-Ir2-Te3 | | 180 | |
| Interlayer | Ir-Ir (Å) | 5.4050 | Ir2-Ir2 | | 5.3133 | Ir2-Ir2 | | 5.3133 | 6.4320 |
| | Te-Te (Å) | 3.5280 | Te1-Te1 | | 2.9384 | Te1-Te1 | | 2.9207 | 2.8921 |
| B$_0$ (GPa) | | 132(9) [13] | 95(16) | | | | | 126(5) [13] |
| ρ = ρ$_0$ +AT$^n$ | | | n≠2 | | | | | n≠2 (Ir$_{0.97}$Te$_2$) n≠2 (Ir$_3$Te$_8$) [14] |
| ρ$_{300K}$ (mΩ cm) | | ~0.3 [11] | 0.12 | | | | | 0.51 (Ir$_{0.97}$Te$_2$) ~0.5 (Ir$_3$Te$_8$) [14] |
| χ$_0$ (emu mol$^{-1}$) | | −0.5×10$^{-4}$ [11] | −1.0×10$^{-4}$ | | | | | −3.5×10$^{-4}$ (Ir$_3$Te$_8$) [14] |
| χ$_d$ (emu mol$^{-1}$) | | -1.75×10$^{-4}$ | -1.75×10$^{-4}$ | | | | | -6.65×10$^{-4}$ (Ir$_3$Te$_8$) |
| χ$_p$ (emu mol$^{-1}$) | | 1.25×10$^{-4}$ | 0.75×10$^{-4}$ | | | | | 3.15×10$^{-4}$ (Ir$_3$Te$_8$) |
| γ (mJ mol$^{-1}$ K$^{-2}$) | | 4 [11] | 1.8 | | | | | 9 (Ir$_{0.97}$Te$_2$) 11 (Ir$_3$Te$_8$) [14] |
| R$_w$ | | 2.27 | 3.03 | | | | | 2.08 (Ir$_3$Te$_8$) |
| θ$_D$ (K) | | 151 [11] | 240 | | | | | 262 (Ir$_{0.93}$Te$_2$) [24] 249 (Ir$_3$Te$_8$) [14] |